\documentclass[conference]{IEEEtran}
\IEEEoverridecommandlockouts
\usepackage{cite}
\usepackage{amsmath,amssymb,amsfonts}
\usepackage{algorithmic}
\usepackage{graphicx}
\usepackage{textcomp}
\usepackage{xcolor}
\usepackage[nolist]{acronym}
\usepackage{paralist}
\def\BibTeX{{\rm B\kern-.05em{\sc i\kern-.025em b}\kern-.08em
    T\kern-.1667em\lower.7ex\hbox{E}\kern-.125emX}}

\begin{document}


\begin{acronym}
\acro{sg}[SG]{smart grid}
\acroplural{sg}[SGs]{smart grids}
\acro{der}[DER]{distributed energy resource}
\acroplural{der}[DERs]{distributed energy resources}
\acro{ict}[ICT]{information and communication technology}
\acro{fdi}[FDI]{false data injection}
\acro{scada}[SCADA]{supervisory control and data acquisition}
\acro{mtu}[MTU]{master terminal unit}
\acroplural{mtu}[MTUs]{master terminal units}
\acro{hmi}[HMI]{human machine interface}
\acro{plc}[PLC]{programmable logic controller}
\acroplural{plc}[PLCs]{programmable logic controllers}
\acro{ied}[IED]{intelligent electronic device}
\acroplural{ied}[IEDs]{intelligent electronic devices}
\acro{rtu}[RTU]{remote terminal unit}
\acroplural{rtu}[RTUs]{remote terminal units}
\acro{iec104}[IEC-104]{IEC 60870-5-104}
\acro{apdu}[APDU]{application protocol data unit}
\acro{apci}[APCI]{application protocol control information}
\acro{asdu}[ASDU]{application service data unit}
\acro{io}[IO]{information object}
\acroplural{io}[IOs]{information objects}
\acro{cot}[COT]{cause of transmission}
\acro{mitm}[MITM]{Man-in-the-Middle}
\acro{fdi}[FDI]{false data injection}
\acro{ids}[IDS]{intrusion detection system}
\acroplural{ids}[IDSs]{intrusion detection systems}
\acro{siem}[SIEM]{security information and event management}
\acro{mv}[MV]{medium voltage}
\acro{lv}[LV]{low voltage}
\acro{cdss}[CDSS]{controllable distribution secondary substation}
\acro{bss}[BSS]{battery storage system}
\acroplural{bss}[BSSs]{battery storage systems}
\acro{pv}[PV]{photovoltaic inverter}
\acro{mp}[MP]{measuring point}
\acroplural{mp}[MPs]{measuring points}
\acro{dsc}[DSC]{dummy SCADA client}
\acro{fcli}[FCLI]{Fronius CL inverter}
\acro{fipi}[FIPI]{Fronius IG+ inverter}
\acro{sii}[SII]{Sunny Island inverter}
\acro{tls}[TLS]{transport layer security}
\acro{actcon}[ActCon]{activation confirmation}
\acro{actterm}[ActTerm]{activation termination}
\acro{rtt}[RTT]{round trip time}
\acro{c2}[C2]{command and control}
\acro{dst}[DST]{Dempster Shafer Theory}
\acro{fcm}[FCM]{fuzzy cognitive map}
\acroplural{fcm}[FCMs]{fuzzy cognitive maps}
\acro{ec}[EC]{event correlator}
\acro{sc}[SC]{strategy correlator}
\acro{ioc}[IoC]{indicator of compromise}
\acroplural{ioc}[IoCs]{indicators of compromise}
\acro{ot}[OT]{operational technology}
\acro{it}[IT]{information technology}
\acro{cvss}[CVSS]{common vulnerability scoring system}
\acro{cve}[CVE]{common vulnerability enumeration}
\acro{ip}[IP]{Internet protocol}
\acro{ems}[EMS]{energy management system}
\acro{hcpn}[HCPN]{hidden-colored-petri net}
\acro{dsr}[DSR]{demand side response}
\acro{pol}[PoL]{Pattern-of-Life}
\acro{ttp}[TTP]{tactic, technique and procedure}
\acroplural{ttp}[TTPs]{tactics, techniques and procedures}
\acro{oscti}[OSCTI]{open-source Cyber Threat Intelligence}
\acro{apt}[APT]{advanced persistent threat}
\end{acronym}

\bstctlcite{IEEEexample:BSTcontrol}

\title{On Holistic Multi-Step Cyberattack Detection via a Graph-based Correlation Approach}

\author{
\IEEEauthorblockN{%
Ömer Sen\IEEEauthorrefmark{1},
Chijioke Eze\IEEEauthorrefmark{2},
Andreas Ulbig\IEEEauthorrefmark{1},
Antonello Monti\IEEEauthorrefmark{2},
}

\IEEEauthorblockA{%
\IEEEauthorrefmark{1}\textit{IAEW, RWTH Aachen Univserity,} Aachen, Germany\\
Email: \{o.sen, a.ulbig\}@iaew.rwth-aachen.de}
\IEEEauthorblockA{%
\IEEEauthorrefmark{2}\textit{ACS, RWTH Aachen University,} Aachen, Germany\\
Email: \{chijioke.eze, amonti\}@eonerc.rwth-aachen.de}
}

\IEEEoverridecommandlockouts

\IEEEpubid{\makebox[\columnwidth]{978-1-6654-3254-2/22/\$31.00~\copyright2022 IEEE \hfill}\hspace{\columnsep}\makebox[\columnwidth]{ }}

\maketitle

\IEEEpubidadjcol

\begin{abstract}
While digitization of distribution grids through information and communications technology brings numerous benefits, it also increases the grid's vulnerability to serious cyber attacks.
Unlike conventional systems, attacks on many industrial control systems such as power grids often occur in multiple stages, with the attacker taking several steps at once to achieve its goal.
Detection mechanisms with situational awareness are needed to detect orchestrated attack steps as part of a coherent attack campaign.
To provide a foundation for detection and prevention of such attacks, this paper addresses the detection of multi-stage cyber attacks with the aid of a graph-based cyber intelligence database and alert correlation approach.
Specifically, we propose an approach to detect multi-stage attacks by leveraging heterogeneous data to form a knowledge base and employ a model-based correlation approach on the generated alerts to identify multi-stage cyber attack sequences taking place in the network.
We investigate the detection quality of the proposed approach by using a case study of a multi-stage cyber attack campaign in a future-orientated power grid pilot.
\end{abstract}


\section{Introduction} \label{sec:introduction}
The \ac{sg} is a gigantic network spanning over a huge geographical area with heterogeneous communication network~\cite{rajendran2019cyber}. It has been undergoing rapid transformation in recent years due to the need to integrate renewable energy sources (distributed over a large geographical area) into the grid.
These transformations are turning the power grid into an intelligent interconnected system~\cite{vijayapriya2011smart}  with a high degree of digitization; thus, blurring the boundary not only between \ac{ot} and \ac{it} but also between the process network and external (for instance public networks such as Internet) network(s)~\cite{siemers2021modern}.
The consequence of this is that the power grid paradigm is now beginning to include increased interaction between stakeholders involved in power generation, distribution, and consumption. An instance is flexibility exchange within a local energy market~\cite{teotia2016local}.
The increased interaction between different actors not only leads to a more complex situation in terms of cyber-physical sovereignty of the grid~\cite{aldabbas2020smart} but also results in an increased threat surface for potential intrusion into the system by unauthorized third parties~\cite{krause2021cybersecurity}.
Mitigation of the new threat surface in the \ac{sg} requires development and integration of a holistic security concept in the \ac{sg}. This includes preventive measures such as security-by-design principles and also detective and reactive measures~\cite{van2020methods}.
Particularly, detective measures that traditionally focus only on events within the communication layer of the system (e.g., traditional \ac{ids})~\cite{chen2014intrusion} must not only consider suspicious and malicious events but also their underlying meaning and implications in the context of the system~\cite{escudero2018process}.
This contextual consideration of information include accurate interpretation of events to reduce false positives~\cite{anton2017question} as well as details that can be used to derive missing information, resulting in reduced false negatives~\cite{sourour2009environmental}.
It has been reported that efficient data collection and analysis are needed to understand the context of a given cyber attack situation. This requires acquisition and utilization of data from communication layer and data relating to the infrastructure, topology, assets, actors, mission, and their interaction to be stored in a knowledge base ~\cite{kure2019cyber}. 

Thus, this paper aims to address the following challenges:
\begin{enumerate}[(i)]
    \item Development of an efficient approach for collection and aggregation of heterogeneous cyber attack-related data from different domains. 
    \item Development of an appropriate attack model to serve as a basis for contextual assessment of security-related data from the offensive security perspective.
    \item Development of a simple approach for correlating heterogeneous security-related data to enable reconstruction of complex cyber attack campaigns within the \ac{sg}. 
\end{enumerate}

In this paper, we propose a framework that leverages heterogeneous cyber intelligence data to detect complex cyber attacks in \ac{sg}.
We summarize key contributions of this paper as follows:
\begin{enumerate}
	\item We present the current state of the art in cyber intelligence and correlation approaches and highlight problems facing existing detection approaches. 
	\item We present a framework that address the problems associated with the present detection approaches.
	\item We present a use case implemented in a real \ac{sg} pilot to demonstrate the detection quality of our framework.
\end{enumerate}

Additionally, the repository hosting the code implementation for the work will be made available to the public as an open-source at the end of the project in 2022. 

The remainder of this paper are as follow. Section~\ref{sec:background} discusses cyber security in \ac{sg}. 
We discuss our proposed framework for multi-setp attack detection in Section~\ref{sec:framework} and present a validation experiment we performed using the proposed framework in Section~\ref{sec:result}. Section~\ref{sec:conclusion} presents the conclusion and possible extensions to the present work.

\section{Cyber Security in Smart Grids} \label{sec:background}
In this section, we provide an overview of the state of the art of cyber security issues in \ac{sg} with respect to the present work. 

\subsection{Cyber Intelligence in Smart Grid} \label{subsec:background_cyberint}
Cyber intelligence gathering is an important task that can help in defending the power grid against cyber attacks. For instance, log-based cyber threat hunting has been employed when handling sophisticated attacks. However, it should be noted that existing approaches require significant manual effort to retrieve relevant attack information and generally overlook rich external sources of threat knowledge such as \ac{oscti}.
To this end,~\cite{gao2021enabling} developed an automated technique for extracting knowledge about threats (indicators of compromise, \acp{ioc}) and their relations from unstructured \ac{oscti} reports. The authors showed how to use the resulting knowledge to aid threat hunting activity.

Further, ~\cite{sree2021artificial} proposes a realistic and comprehensive model for detecting attacks.
The method splits attack detection process into six stages: aim, scale, equipment, planning, execution and input, and validated the model using Ukrainian electricity grid attacks.
The authors showed how advanced persistent graphs can be used to extract \ac{ioc} from the attacker’s \ac{ttp} for threat hunting purposes.
The method provides a high-level representation of an attack campaign and enables quick assessment of attack progression from both attacker and defender’s perspectives.
In the same vein,~\cite{zongxun2021construction} applied natural language processing model (BERT-BiLSTM-CRF) to automatically extract threat actions and generate \acp{ttp} from advanced persistent threat reports.

Similar to the above mentioned works (particularly, ~\cite{sree2021artificial} and ~\cite{zongxun2021construction}), the present work includes a method for assessing risk and detection of attack actions. It is based on a graph-structured data model (knowledge graph) of the framework which integrates both static and dynamic network data to enable capabilities such as fast and dynamic generation of attack graphs for the network. The attack graph forms the basis for further analysis of attack progression. More details on this are presented in Section~\ref{sec:framework}.

\subsection{Contextual Detection of Cyber Attacks} \label{subsec:background_detection}
\ac{ids} has been the de facto tool for detecting attacks in networks and networked systems. An \ac{ids} can be used to monitor the \ac{ict} network or hosts attached to it for attack-related activities such as login attempts, network scans, suspicious log traffic or syslog without considering process semantics of the power grid~\cite{chromik2017context}.
Generally, modern \ac{ids} achieves uses either supervised machine learning approach (misuse approach) that employs attack signatures or unsupervised machine learning approach (anomaly based approach) to automatically identify indicators of attack.
A combination of the two approaches has also been proposed for improved detection capability of the \ac{ids}~\cite{aljamal2019hybrid}. 
While the above approaches perform generally well for single-stage attacks in most cases, they have been shown to be ineffective against multi-stage attacks.
In such cases, contextual information can be used to used improve performance and ensure fast detection capability.

Contextual detection involves aggregation of data (for instance, logs and event data) from various sources (e.g. \ac{ids}, firewalls, etc), normalization of the data to a common format and finally synchronization of associated event fields (e.g. timestamps) for further data processing and for performing alert correlation~\cite{bryant2017novel, radoglou2021spear}.
Apart from data gathering, attack modeling is also important for contextual detection. A common way of modeling a multi-stage attacks is by creating its attack graph representation. 
 It enables extraction of contextual information via correlation and reasoning over the attack-related data to classify or infer characteristics and relations between entities involved in the attack~\cite{sen2021towards}.

An attack graph captures two concepts: successor and predecessor nodes, and helps to discover different paths an attacker unleashing a multi-step attack can take to reach the target~\cite{angelini2018attack}.

Apart from attack graph representation, other approaches have also been proposed in the research community for extracting contextual information when handling multi-stage attacks.
For instance,~\cite{aparicio2018multi}  employ contextual information in the form of \ac{pol}, and expert knowledge on true network behavior to develop a novel \ac{ids} that can detect multi-stage attack in real-time. One of the benefits of this approach is that it does not require prior training.
In addition to use of \ac{pol}, \cite{aparicio2019addressing} propose an anomaly-based \ac{ids} that improves the efficiency of the \ac{ids} by using \ac{dst} theory and \acp{fcm} to encode contextual information. To improve performance and robustness to missing alerts/detections, the present work uses \ac{hcpn} models~\cite{yu2007improving} to identify the underlying attack behind set of alerts generated by attack detection systems.

\subsection{Problem Analysis} \label{subsec:background_problem}
Although several techniques and frameworks have been proposed to tackle cyber attacks against power grids, it is worthy of note that there are some challenges that are drastically affecting their performance~\cite{rajendran2019cyber}.
One major challenge is that of gaining access to relevant cyber attack data, availability of resources to quickly process the data and skills to do so.
Other key challenges facing cyber attack detection and mitigation in \acp{sg} include:
\begin{itemize}
    \item Lack of explainability of machine/deep learning models used in anomaly based \acp{ids}
    \item High false positives in detection results, especially in anomaly-based \acp{ids}
    \item Lack of flexibility to changing threat landscape. This is the major issue with signature-based \acp{ids}
    \item Lack of overarching context regarding the attack objective for reactive countermeasures
\end{itemize}

The specific property of \ac{sg} enables deterministic and model-based approaches that exploit domain-specific knowledge.
In particular, \ac{ot} in \ac{sg} is characterized by the following:
\begin{inparaenum}[i)]
    \item the predominance of machine-to-machine communication (deterministic communication behavior),
    \item homogeneity of assets (simplification of modeling),
    \item specific network requirements for different device types (accurate network modeling),
    \item hierarchical control structures (plausibility check),
    \item long lifetime of installed field devices (static infrastructure),
    \item pre-authorization of field devices against local servers (simple authentication process), and
    \item defined life cycle management and maintenance processes (remote access processes).
\end{inparaenum}
These properties enable a more straightforward model-based detection approach where domain-specific knowledge can be structured in a given ontology model.
Thus, in this paper, we propose a domain-specific knowledge-based system that can be used for multi-stage cyber attack detection.

\section{Multi-Staged Attack Detection System} \label{sec:framework}
In this section, we present our proposed multi-stage attack detection framework and the method we use in cyber-intelligence acquisition and alert correlation.

\subsection{Framework Overview} \label{subsec:framework_overview}
\begin{figure}
    \centerline{\includegraphics[width=\columnwidth]{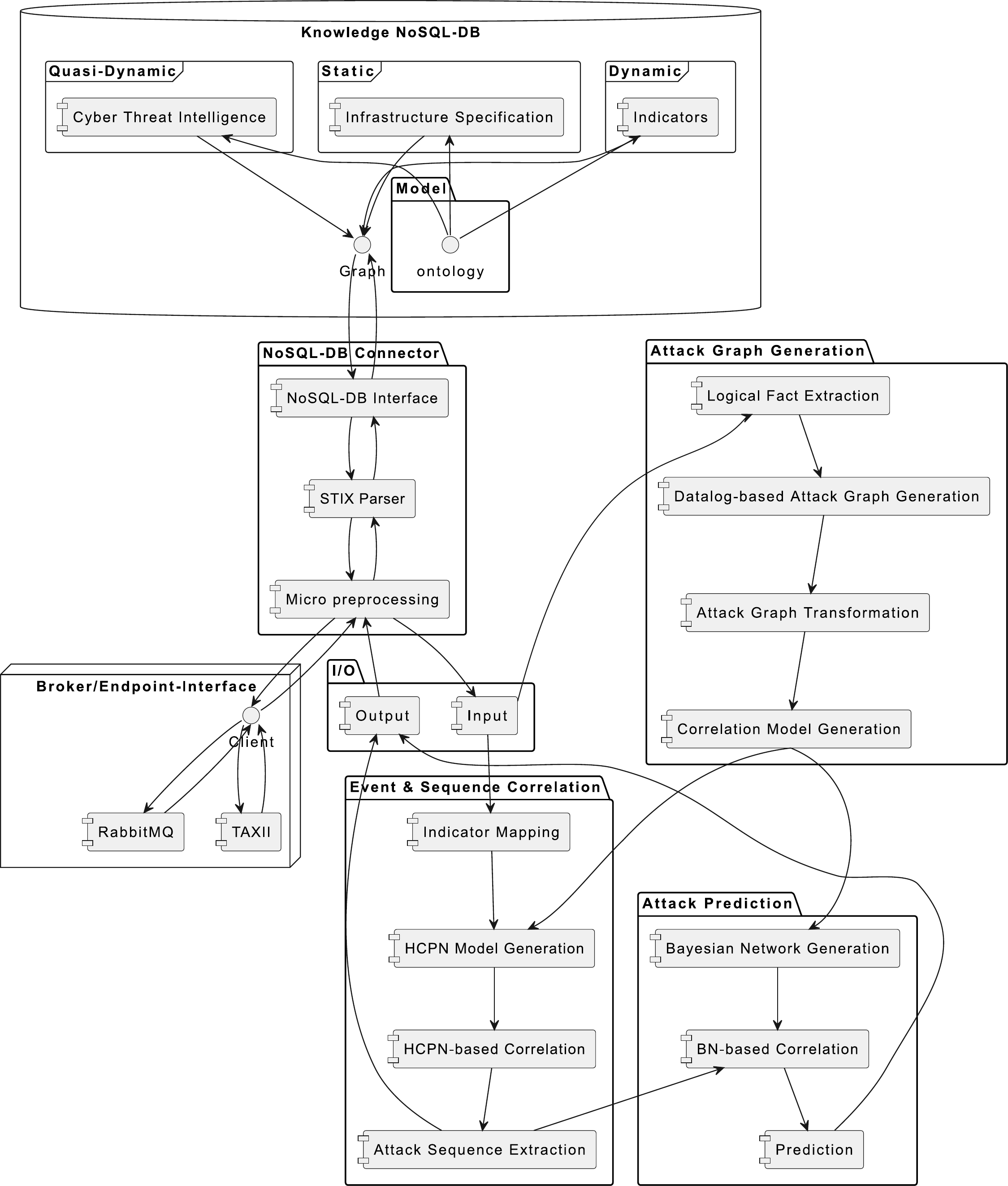}}
    \caption{Structural overview of the framework architecture with the core components of the knowledge base and attack correlation.}
    \label{fig:cosim_framework}
    \vspace{-1em}
\end{figure}
Figure~\ref{fig:cosim_framework} illustrates the architecture of the proposed framework, which at its core realizes a knowledge-based approach to correlate and predict attack sequences.
We modeled a cross-domain descriptive ontology model in RDF to improve and enrich the knowledge management process. The ontology model combines infrastructure-specific data with cyber threat information as well as dynamically retrieved indicators in a coherent and comprehensive graph-based data model.
This approach helps to enrich available data and knowledge through linkage and classification as they are collected and stored.

For the connection to the knowledge base, we specifically developed a graph-based database connector module in Python that handles all access processes. It manages data exchange between the access-requesting module(s) and the database, handles all intervening micro-managements such as format conformance transformation and parsing as well as pre-processing.
Additionally, the connector is also the main access point for external components requesting access to our database via a broker messaging system (e.g., RabbitMQ) or a direct client-server endpoint connection.
Within the framework, we use the connector to retrieve necessary data and specifications for attack and event correlation, and prediction processes.

The first step for these processes is the extraction and generation of a logic-based attack graph via the attack graph generation module using Datalog.
In this module, we implemented an approach to extract factual data such as infrastructure information and cyber threat data from the knowledge base and generate the attack graph for the given attack scenario.
This approach enables generation of a consistent and coherent attack graph based on currently observed and available data in the knowledge base.

To better estimate the likelihood that an attack has occurred, we convert the attack graph into a correlation-enabled model in Python using NetworkX, which is then provided to the attack sequence correlation module.
Until this step, all processes and computations such as knowledge base creation with (quasi-) static (infrastructure-related) information and attack graph generation and conversion are performed in larger time domains.
Because the information needed for attack graph generation is unlikely to change frequently, performance-intensive tasks can be distributed into defined time ranges that reflect the frequency of infrastructure changes.
Thus, a periodic update process can be implemented for the knowledge base. Updates are performed either when changes occur within the infrastructure or at defined time intervals.
The updates are performed through the interconnector, which translates the structured knowledge into queries according to the ontology.

To provide a basis for deriving appropriate remediation measures, we use the attack sequence in conjunction with a prediction model generated from the attack graph to forecast possible next attack steps.

\subsection{Knowledge Base} \label{subsec:framework_knowledgebase}
\begin{figure}
    \centerline{\includegraphics[width=\columnwidth]{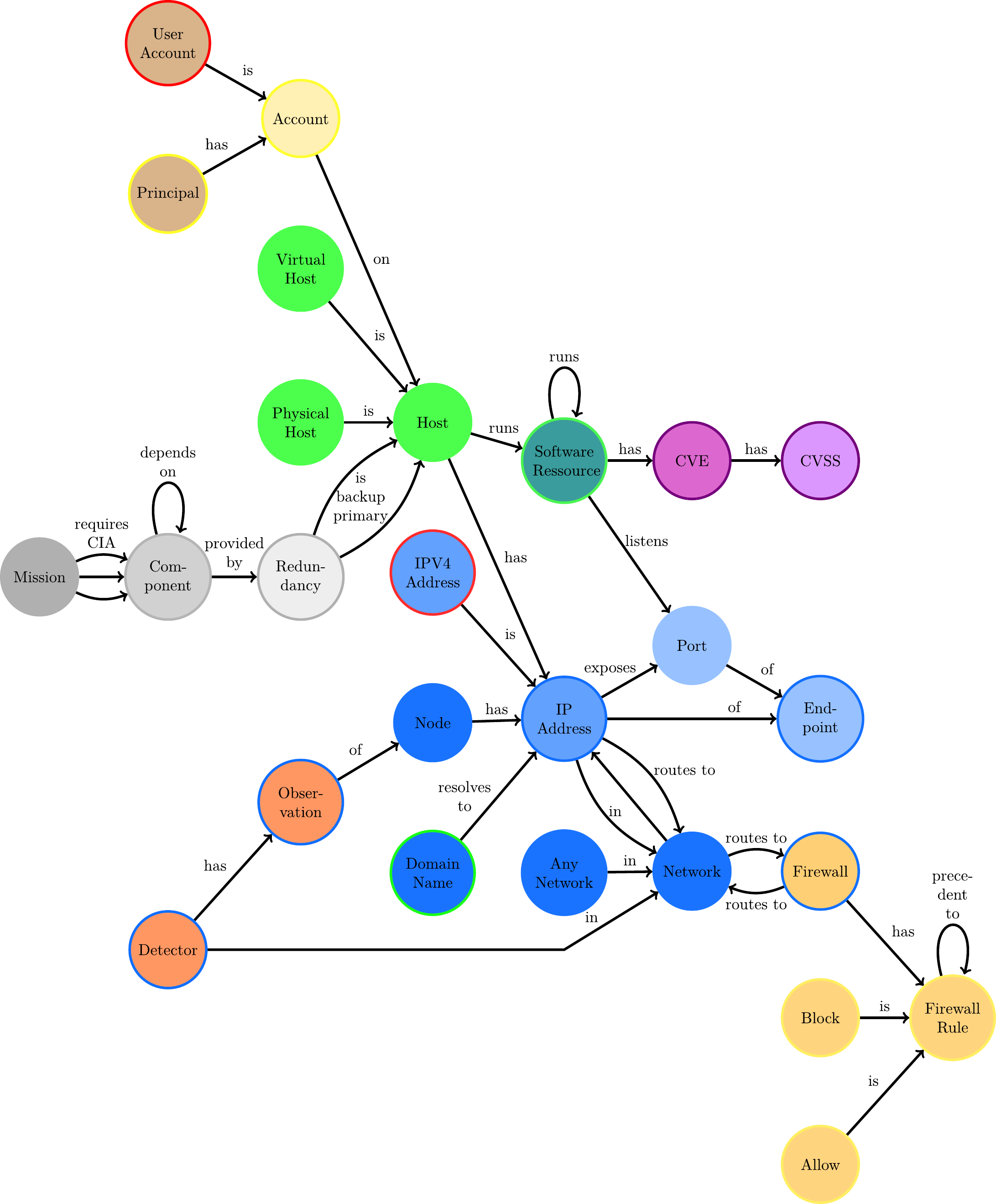}}
    \caption{Structural overview of the ontology model for merging heterogeneous data into a consistent cyber intelligence database.}
    \label{fig:cosim_ontology}
    \vspace{-1em}
\end{figure}
The knowledge base aims to transform heterogeneous data into a consistent graph-based data model which is implemented as a No-SQL graph database using Neo4j~\cite{miller2013graph}.
This allows us to specify and compile the knowledge of the system including both static information such as the system topology and dynamic data such as attack indicators into the knowledge base.
In this way, we can easily adopt new use cases and increase situational awareness by adding context to the received event messages and indicators.

In the following, we describe the current ontology using class views.
The designed ontology model is shown in Figure~\ref{fig:cosim_ontology}.
All nodes represent classes with all their possible relationships (edges) to other classes.
They do not represent individual class instances (objects).

The blue nodes represent the network layer.
Further, we define the \ac{ip} networks that are present in the infrastructure.
The \ac{ip} address nodes are connected to the network nodes to indicate their membership.
We also define routing nodes that can connect different subnets.
Often, these also have firewall functions.
We define routing between hosts and networks using the route relationship.
The firewall, which is part of the yellow authorization layer, can have firewall rules: allow and block rules.
These are used to describe various rules that determine whether or not certain communications traffic such as \ac{ip} packets will be forwarded by the firewall.
This information is important when creating attack graphs as it determines which host can access a given host and how.
The other part of the privilege layer includes user accounts and principles which are used to specify different accounts and their privileges on each host.

The host layer (green) specifies which hosts exist in the infrastructure.
Hosts can be assigned multiple \ac{ip} addresses and thus be part of multiple networks.
Software resources run on them, which in turn can use other software resources.
Software resources may have ports associated with them if they are running as network services accessible through a network interface.
This information is extremely important because it shows which software resources are remotely accessible and can potentially be the target of an attack vector.
Further, our ontology also specifies known vulnerabilities of various software resources and their properties.
For this purpose, we use the \ac{cve} and the \ac{cvss}, which are associated with the corresponding software resources and are shown in purple.

The gray nodes represent the mission layer.
The mission layer defines the critical functions provided by the infrastructure, i.e., the potential attack targets of an attacker.
For example, the operational goal of a control system is defined by the mission layer, which for an \ac{ems} can be, for example, optimization of self-consumption, which is what the attacker potentially might try to interfere with.

In general, the knowledge base helps in modeling of the dependencies and interactions between components present in the system.
Further, we associate dynamic information such as alerts are with network nodes that are involved in the observation.
Each observation belongs to a detector that triggered the observation.

\subsection{Alert Correlator} \label{subsec:framework_alertcorrelation}
This section describes the method we employ to correlate observations with data in the knowledge base.
The correlation method is based on Petri net model~\cite{peterson1977petri}, a mathematical modeling language typically used to describe distributed, discrete event systems and to model concurrent behavior.
In the proposed approach, we employ Petri net to model cyber threats, detect multi-stage attacks, and to perform event correlation.
In particular, we use \ac{hcpn} models~\cite{yu2007improving} because it offers richer representation than ordinary Petri net.

We use ``places'' in the \ac{hcpn} to represent security conditions and targets that describe the attack process while transitions represent the atomic attack steps executed by the attacker.
As an extension to regular vanilla Petri nets, an \ac{hcpn} also has outgoing arcs from a transition to observation nodes that can be triggered when the connected transition fires.
This models the situation where an attacker's action can trigger indicators or alarms.
The reason for this is that the defender cannot observe the attacker's actions directly but only through the triggered indicators. Further, these indicators can sometimes be ambiguous or uncertain due to imprecise information provided by the sensors or the extracted security sources.
The Petri net model itself, a bipartite graph, can be directly derived from the logical attack graph we generated based on our infrastructure knowledge in addition to the cyber threat information.
Particularly, we use MulVal~\cite{ou2005mulval} to generate attack graphs and in performing risk assessment~\cite{frigault2008measuring}.
The resulting risk values from the assessment are used to calculate the probability of attack actions in terms of offensive security and to evaluate the costs and benefits of attack actions from the attacker's perspective.
Thus, the concept of attack action probability in our work is understood from a decision-theoretic perspective on whether the attack action makes sense for the attacker to perform.
However, this concept depends on prior knowledge about the attacker's main objective, which is also an essential part of attack graph generation.
This is why our attack graphs contain several main attack objectives (nodes without outgoing edges).  Formally, we define attack action probability, $\mathrm{Pr}$ as
\begin{equation}
	\mathrm{Pr}=\mathrm{p}\left(A_a \mid A_{\mathrm{g}}, \mathrm{precons}\right)
\end{equation}
where $\mathrm{A}_{\mathrm{a}}$ denotes attack action, $\mathrm{A}_{\mathrm{g}}$ denotes attack goal and precons denotes preconditions for the attack action.

Similarly, all connections between transitions, places, and observations are assigned probability-based confidence values that indicate the likelihood of the node at the outgoing end given the node at the incoming end.
In this way, we can represent the uncertainty in the attacker's actions, their preconditions, their success rate, and the false-positive and true-positive rates.
We also define a location and transition corresponding to harmless actions falsely issued as alerts in the system.
This facilitates mapping of alerts that are not strongly correlated with attack steps to harmless false-positive activities in the system.
The probabilistic nature of our model allows us to specify various probabilistic computations according to~\cite{yu2007improving}.
Thus, a dynamic programming-based approach can be taken that allows us to derive a sequence of attacker actions from a sequence of alerts.
We use this approach for event correlation as it requires limited assumptions and computational complexity of $O(m^2 n)$, where m denotes the number of transitions and n denotes the number of observations.
In the end, we obtain the sequence of attack actions that maximizes the likelihood of the model.

The resulting attack sequence contains a more strict path of actions that represents the attack evolution.
The \ac{hcpn} model represents possible attack actions on the infrastructure based on the knowledge given by the generation timestamp.
As a consequence, the flexibility of the model is tied to the knowledge base. The model is adapted to changes in the knowledge base by generating an updated model.
By combining the model with a probability-based prediction model such as Bayesian network~\cite{kondakci2010network}, likely next attack steps are predicted.
In this way, we detect historical evolution of attacks based on existing observations and predict potential future actions by using Bayesian network.

Because of this situation-aware detection approach, our approach can reduce false positives and false negatives rates by contextually evaluating the attack indicators within the attacker model.
Thus, the attack is evaluated in terms of whether the indicators are plausible in their chronology and also with respect to the trajectory path of the attacker model.
The drawback of the proposed approach is the increased information required for its deployment and operation.
Situational awareness in our approach requires detailed information about the infrastructure, devices, networking, and configuration, which in a less detailed state may affect the quality of detection obtained by using our framework.

\section{Case Study \& Discussion} \label{sec:result}
In this section, we investigate the detection quality of the proposed framework through a case study.

\subsection{Case Study Scenario} \label{subsec:result_proecdure}
In this case study, we demonstrate the proposed framework in a scenario mainly set in a web-based technology environment where a central platform acts as a \ac{dsr} platform and performs self-consumption optimization for a microgrid.
The microgrid consists of household loads with \ac{pv} and \ac{bss}.
The households are connected to the \ac{dsr} platform via smart meters and provide their metering data while the \acp{der} are controlled via photovoltaic inverters.
In addition, the \ac{dsr} platform also has access to the Internet via a firewall.
Further, another firewall also provides access to the process network containing the control components.

Within the infrastructure, the attack scenario is defined as a multi-stage attack campaign that attempts to penetrate the \ac{dsr} network from the Internet to compromise the platform and cause power imbalance by manipulating process data.
Specifically, the attack exploits vulnerabilities in the SSH network service (enumeration of valid usernames) and initiates a brute force attack to gain access to the \ac{dsr} platform via SSH (dictionary attack).
After successfully gaining access to the \ac{dsr} platform, the attacker performs activities to elevate his privileges and also to gain write privileges on the platform (e.g., SUID-based privilege escalation).
After gaining sufficient privileges on the \ac{dsr} platform, the attacker manipulates the smart meter data received from the \ac{dsr} platform and attempts to create a false grid state that suggests a low-load situation in which the energy supply is provided by the \ac{dsr} and \ac{bss} components must be reduced.
The \ac{ems} logic of the \ac{dsr} platform concludes that the power supplied by the \ac{pv} and \ac{bss} components must be reduced and sends the corresponding control commands.
However, these control commands are sent as grid-state harmful commands, i.e., false commands that compromise grid stability by causing a significant power imbalance.
At this point, the attacker has achieved its global objective.
Detector components are deployed within the infrastructure to monitor communication behavior and process data consistency to provide attack indicators.

\subsection{Case Study Results} \label{subsec:result_res}
\begin{figure}
    \centerline{\includegraphics[width=\columnwidth]{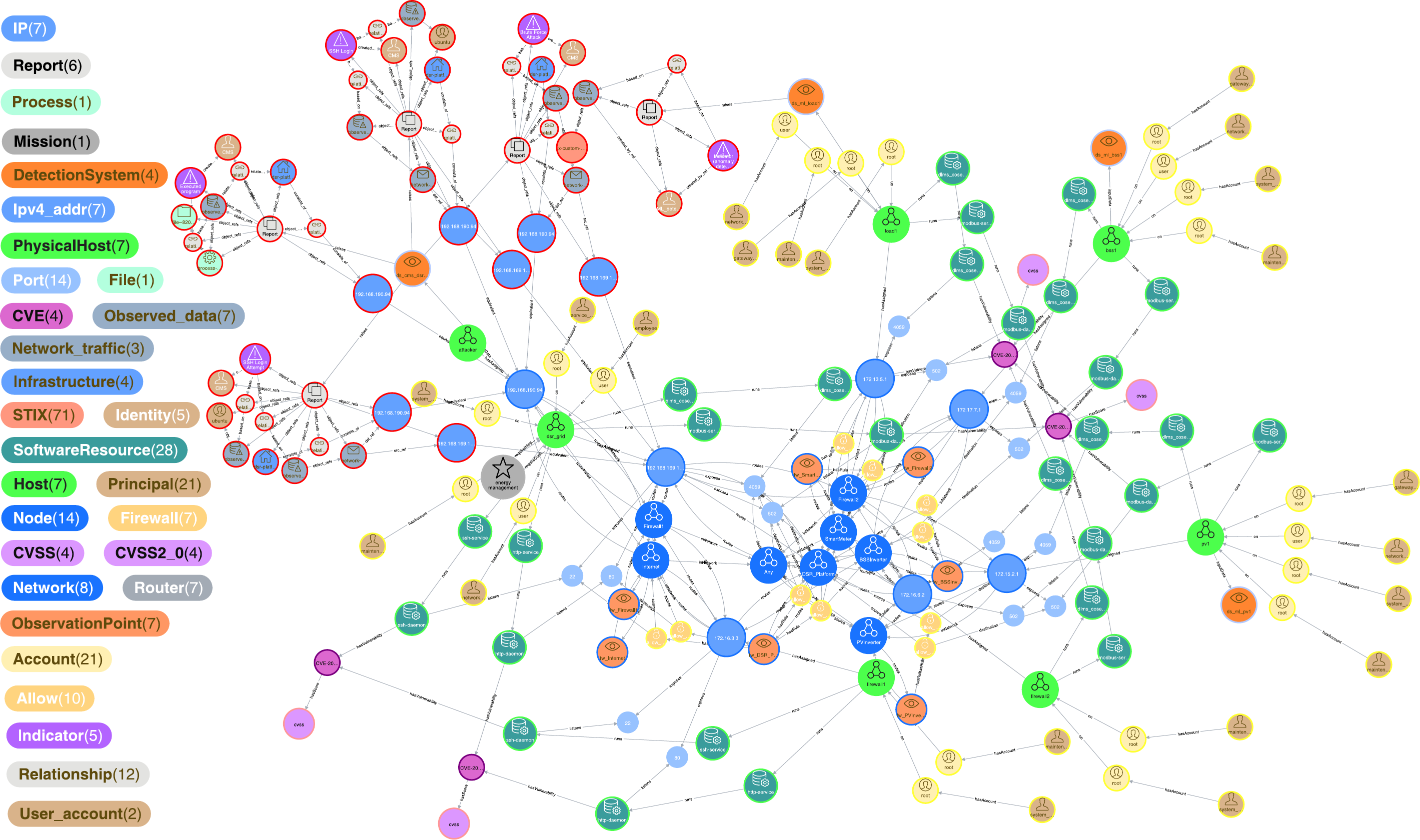}}
    \caption{Illustration of the knowledge base graph created for this case study.}
    \label{fig:result_knowledgebase}
    \vspace{-1em}
\end{figure}
Based on the case study scenario, the knowledge base is populated (cf. Figure~\ref{fig:result_knowledgebase}).
Particularly, the infrastructure information such as the facilities, assets and network components were specified according to our ontology model.
Semantic information such as firewall rules and detector placement are also included.
During the case study scenario execution, the observations reported by the detectors are added to the database.
Based on this information, facts are extracted and used to create the attack graph to determine possible attack actions.

\begin{figure}
    \centerline{\includegraphics[width=\columnwidth]{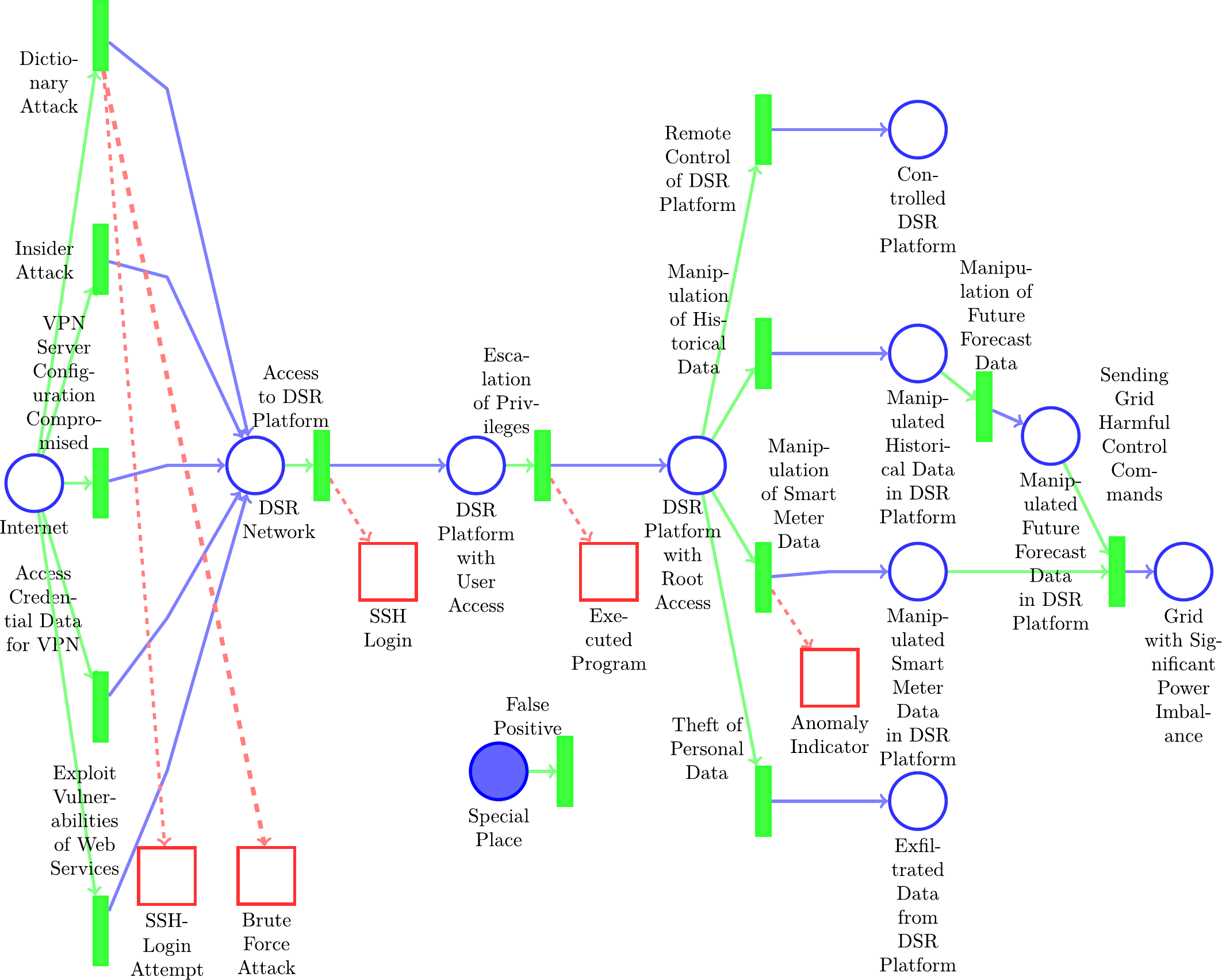}}
    \caption{Illustration of the resulting \ac{hcpn} model for the case study. The blue colored places represent the resources gained after the attack (pre-condition), the green transition represents the actual attack actions (post-condition), while the red nodes represent the observation (alert).}
    \label{fig:result_petrienet}
    \vspace{-1em}
\end{figure}
Figure~\ref{fig:result_petrienet} illustrates the resulting \ac{hcpn} model generated based on the knowledge graph and the obtained attack indicators.
Within the generated \ac{hcpn} model, the dictionary attack action is associated with the SSH login attempt and bruteforce attack indicators.
Further, the \ac{dsr} network access action is linked to the SSH login indicator, while the privilege escalation attack is associated with the executed program indicator.
Smart meter data tampering is detected as an anomaly indicator.

Based on the \ac{hcpn} model, the alert correlation process generates a sequence in the model to which corresponding observations are mapped.
Table~\ref{tab:result_correlation} shows the result of the correlation process, where $P_r$ represents the risk probability that the corresponding transition node is triggered, while $A_a$ represents whether the transition node is actually triggered.
The $A_p$ attribute here represents the predicted attacker's next steps.
Because the attacker in our case study followed a strict action sequence to cause an energy imbalance in the network, the correlated sequence corresponds to the path leading to this goal based on supporting observations.

\begin{table}[h!]
\vspace{-2em}
\centering\small
\caption{Result of the correlation process in the case study.}
\begin{tabular}{||p{5.5cm} p{0.5cm} p{0.5cm} p{0.5cm}||}
 \hline
 Attack Step & $P_r$ & $A_a$  & $A_p$\\
 \hline\hline
 Exploit Vulnerabilities of Web Services & 0.3 & - & -\\
 Access Credential Data for VPN & 0.53 & - & -\\
 VPN Server Configuration Compromised & 0.13 & -& -\\
 Insider Attack & 0.33 & - & -\\
 Dictionary Attack & 0.3 & X & -\\
 Access to \ac{dsr} Platform & 0.26 & X & -\\
 Escalation of Privileges & 0.25 & X & -\\
 Remote Control of \ac{dsr} Platform & 0.61 & - & -\\
 Theft of Personal Data & 0.61 & - & -\\
 Manipulation of Smart Meter Data & 0.96 & X & -\\
 Manipulation of Historical Data & 0.24 & - & - \\
 Manipulation of Future Forecast Data & 0.06 & - & -\\
 Sending Grid Harmful Control Commands & 0.93 & - & X\\
 \hline
\end{tabular}
\label{tab:result_correlation}
\end{table}

Thus, the result of the correlation process fully corresponds to the actual attack path in which the attacker gained access to the \ac{dsr} network through the dictionary attack, elevated its privileges on the \ac{dsr} platform, and manipulated the smart meter data to cause a power imbalance.
However, because there are no other observations about the power imbalance caused due to the attack, the last steps of the \ac{hcpn} are not shown as activated.
However, if we access the attribute $P_r$, the probability or risk of activating the steps, we can see that the last attack step ``sending grid harmful control commands'' has a relatively high probability value, indicating a very likely evolution to this branch of the model.
Accordingly, the $A_p$ attribute for this step indicates a likely next step of the attack that will occur according to the correlated attack sequence.
In the context of the case study, the detected attack result of our approach corresponds to the actual course of the attack and the planned target.
Both the historical and future actions were correctly detected and reported.

Furthermore, in this case study, our approach performed as follows:
\begin{inparaenum}[i)]
 \item population of the knowledge base took $4s$ for 232 nodes,
 \item generation of the attack graph took $8s$ for 860 facts,
 \item attack emulation itself took $229s$ for 4 attack steps,
 \item attack correlation took $3s$ for 198 nodes, and
 \item it took $4s$ to generate and transmit the report.
\end{inparaenum}
Considering that the last two steps in the detection are performed frequently, a detection time of less than $10s$ is a reasonable time frame.
Thus, based on the correlation process, the attack campaign can be reconstructed in a reasonable time span. The next steps to be taken by the attacker can also be predicted.

\subsection{Discussion} \label{subsec:result_dis}
Our proposed framework can reliably reconstruct the attack campaign and even predict the next steps of the attack.
However, to accurately reflect the actual situation, a sufficient number of observations is required.

The alert correlation approach is highly dependent on these observations and thus on the underlying placement of detectors.
If there are not enough detectors to monitor the system, the bias of the correlation process is based more on the logic-based risk assessment, where the probability of easily achievable steps and objectives influences the conclusion.
This can become an issue in attack strategies with multiple targets when there are multiple objectives rather than just one.
A high level of observation based on a large number of detectors can reduce this ambiguity to a set of specific objectives.
Thus, if there are not enough observations available, this correlation and prediction process will favor the objectives that are easy to reach from an attack perspective.

Additionally, the level of information in the knowledge base, particularly the specification of the infrastructure, plays a critical role in the quality of detection.
A detailed description of the underlying infrastructure and network architecture helps in accurate generation of the attack graph and thus influences the correlation and prediction methods.
Missing information can lead to the creation of inaccurate attack graphs that describe a different attack model and thus a different perception of the current situation.
Consequently, the attack correlation and prediction concludes to a different development model of the attack, which results to a higher number of false positives or false negatives.

Therefore, the infrastructure specification and corresponding cyber threat information need to be as accurate and complete as possible to achieve optimal correlation and prediction results.
Nevertheless, in this case study investigation, we demonstrated that our proposed framework can reliably detect multi-stage cyber attacks and predict their next steps, provide high detector coverage and complete information base.

\section{Conclusion} \label{sec:conclusion}
The emerging cyber threat situation for critical infrastructures, especially for power grids, requires a holistic security approach consisting not only of preventive but also detective and reactive measures.
To achieve this goal, we propose an approach capable of reconstructing complex cyber attack campaigns.
The proposed approach is based on knowledge base that fuses heterogeneous data into a consistent graph-based data model.
The knowledge base is used to detect multi-stage cyber attacks by performing \ac{hcpn}-based alert correlation and attack step prediction using a Bayesian network.
We validated the detection quality of our framework using a case study and showed that it can reliably reconstruct multi-stage cyber attacks.
However, the detection quality of the  framework depends on the detector coverage in the system and the completeness of the knowledge base.
a possible future work would be to address the information dependency issue of the framework by developing a robust correlation approach that leverages the graph structure of the knowledge base in a multi-layer correlation approach.
For instance, similarity correlation can be performed based on component attributes and the topological relationship described in the knowledge graph to identify other potential components of interest.

\noindent\textsc{Acknowledgments}\hspace{1em}
This work is partly funded by the European Union’s Horizon 2020 research and Innovation programme under grant agreement N°832989.

\bibliographystyle{IEEEtran}
\bibliography{conference_101719}

\end{document}